\documentclass[prb,onecolumn,showpacs,preprintnumbers,amsmath,amssymb,assume]{revtex4}
\usepackage{graphicx}				
\usepackage{dcolumn}				
\usepackage{bm}						
\newcommand{\Bs}{&&\hspace{-3.2mm}} 

\begin{document}
	
	\title{Seismic characterization of multiple fracture sets at Rulison Field, Colorado}
	
	\author{Ivan Vasconcelos$^1$ and Vladimir Grechka$^{2}$}
	\affiliation{$^{1}$Colorado School of Mines, currently at Utrecht University, The Netherlands \\
		$^{2}$Shell International Exploration and Production, currently at Marathon Oil Company}
	\date{\today}
	
\begin{abstract}

Conventional fracture-characterization methods assume the presence of a single set of aligned,
vertical cracks in the subsurface. We relax this assumption and demonstrate the feasibility of
seismic characterization of {multiple} fracture sets. Our technique relies on recent numerical
findings, indicating that multiple, differently oriented, possibly intersecting planar cracks
embedded in an otherwise isotropic host rock result in a nearly orthorhombic (or orthotropic)
effective medium. Here, the governing parameters of crack-induced orthotropy are estimated from 3D,
wide-azimuth, multicomponent seismic reflection data acquired over the tight-gas Rulison Field in
Colorado, USA.

We translate strong azimuthal variations of the normal-moveout velocities into interval crack
densities, fracture orientations, type of fluid infill, and velocities of the P- and S-waves in an
unfractured rock. Our inversion procedure identifies a set of cracks aligned in approximately
WNW-ESE direction in the western part of the study area and multiple, likely intersecting fractures
in its eastern part. We validate both our underlying theoretical model and the obtained estimates
by two independent measurements: (i) the estimated fluid-infill parameter indicates dry cracks as
expected for the gas-producing sandstones at Rulison; and (ii) the obtained crack orientations are
supported by measurements in a well. As a by-product of fracture characterization, we build an
anisotropic velocity model of the Rulison reservoir that, we believe, is the first orthorhombic
velocity field constructed from surface seismic data.

\end{abstract}
\maketitle

\section{Introduction} \label{sec:intro}

Naturally fractured rocks are widely recognized as potential hydrocarbon-bearing formations.
Perhaps this is why the 2004 Summer Research Workshop of the Society of Exploration Geophysics
(SEG) and the European Association of Geoscientists and Engineers (EAGE) was held to determine the
state of the art of seismic characterization of fractured reservoirs. The workshop demonstrated
that current industry capabilities of inverting parameters of small, sub-seismic fractures are
limited to a single set of penny-shaped, vertical cracks embedded in an otherwise isotropic host
rock. This arrangement leads to the effective horizontal transverse isotropy (HTI), its
anisotropy coefficients can be estimated from seismic data and related back to the fractures. The
papers presented at the workshop exploited HTI model and discussed various approaches to inferring
the crack density, fracture orientation, and the type of fluid infill from such seismic signatures
as normal-moveout (NMO) velocities\cite{Berthetetal2004}, amplitude-versus-offset (AVO) variations\cite{delMonteetal2004, GrayTodorovic2004, Minsleyetal2004, Todorovicetal2004}, and shear-wave splitting coefficients\cite{GaiserVanDok2004}.

As all of the above mentioned techniques are inherently based on measuring the magnitude of azimuthal
anisotropy in a selected seismic signature and translating it into the parameters of a single set
of cracks, they are bound to produce misleading results in the presence of differently oriented
fractures. Such fractures are observed at numerous outcrops throughout the globe and usually
identified in the borehole televiewer data\cite{Laubachetal2004}. Given the ubiquity of
multiple fracturing, the inadequacy of existing fracture-characterization technology might be
explained by an inherent nonuniqueness in inverting the effective anisotropy (estimated from
seismic data) for the parameters associated with multiple crack systems. Indeed, the number of
fracture-related parameters can be arbitrarily large, whereas the maximum number of independent
effective stiffness coefficients for any homogeneous rock volume is 21. While the number~21
obviously imposes an ultimate limit on the number $N$ of fracture sets that can be uniquely
resolved from long-wavelength seismic data, this~$N$ is typically greater than one. Grechka and
Tsvankin\cite{GrechkaTsvankin2003} used the linear slip theory\cite{Schoenberg1980} to derive limits for~$N$. They found, in agreement with earlier results\cite{BakulinetalPart22000}, that seismic data can
uniquely constrain the parameters of up to two vertical, rotationally-invariant fracture sets
embedded in isotropic background rock.

To make further progress in fracture characterization, one has to rely on certain microstructural
information related to the cracks. Such information was utilized in the effective media theory
proposed by Kachanov\cite{Kachanov1980}. Being similar to the popular linear slip theory\cite{Schoenberg1980} for dry cracks, Kachanov's theory adds the following important insight: the collective contribution of
multiple sets of dry fractures to the effective elasticity is largely controlled by a symmetric,
second-rank crack-density tensor; such fractures are sometimes termed scalar\cite{SchoenbergSayers1995}.
The ability of representing multiple crack systems in an otherwise isotropic rock by a
{second-rank} tensor leads to {orthorhombic} effective symmetry for {any}
orientations of multiple fracture sets, implying that arbitrarily oriented, vertical cracks can
be replaced by {two} mutually {orthogonal} sets, called the principal sets, as
far as propagation of long seismic waves is concerned. (Here and throughout the paper, the term
``long'' refers to waves whose lengths are orders of magnitude greater than the crack sizes.) The
effective orthotropy induced by dry cracks residing in a purely isotropic host rock turns out to be
remarkably simple: it is fully described by only four independent quantities instead of nine needed
for general orthotropy. These quantities are the two principal crack densities and two Lam\'{e}
constants of the isotropic background.

Those findings\cite{Kachanov1980, Kachanov1993} are extremely valuable for fracture 
characterization because they reduce any number of vertical fracture sets to just two and,
thus, provide a theoretical basis for estimating parameters of multiple systems of
fractures. Prior to using these theoretical predictions for parameter estimation,
however, it is important to establish their accuracy and the range of validity. This
work has been carried out in a series of papers\cite{GrechkaKachanovORTfrac2006, GrechkaKachanovXfrac2006, GrechkaetalSfrac2006}. Their authors performed finite element simulations of effective elasticity for a wide variety of crack arrays that contained planar, open and partially closed, intersecting and
non-intersecting, circular and irregular dry fractures. The results\cite{GrechkaKachanovORTfrac2006, GrechkaKachanovXfrac2006, GrechkaetalSfrac2006} demonstrate that 
deviations from effective orthotropy do not exceed 2\% in the entire range of the crack
densities expected in naturally fractured formations. Moreover, the effective orthotropy holds for liquid-filled cracks even with a better accuracy than for dry cracks because of the stiffening of fractures by fluids and a subsequent
reduction of crack contribution to the overall elasticity\cite{GrechkaKachanovORTfrac2006}. Following numerical
verification of the effective crack-induced orthotropy, a technique for characterization of multiple vertical fracture sets was proposed\cite{GrechkaKachanovORTfrac2006}. While being
similar to an earlier suggested method\cite{BakulinetalPart22000}, the former does not require
information about the vertical velocities. It utilizes only multiazimuth, multicomponent
seismic reflection data to estimate the orientations and crack densities of
two principal fracture sets, the density-normalized Lam\'{e} constants of the background,
and the so-called fluid factor that varies from 0 (when the cracks are dry) to~1 (when the
bulk modulus of the infill approaches that of the background).

Here we apply this fracture-characterization technique to a 3D, 9C, wide-azimuth data set acquired
by the Reservoir Characterization Project (Colorado School of Mines) over the tight-gas Rulison
Field located in Colorado, USA. We invert the interval NMO ellipses of the P, S$_1$ (fast shear), and S$_2$
(slow shear) waves for the two principal crack densities, fracture azimuths, and fluid factors. The
fracture orientations obtained from seismic are supported by direct borehole measurements, whereas
the fluid factors turn out to be less than 0.01, as expected for the gas-bearing sandstones at the
Rulison. Overall, we find the model of crack-induced orthotropy to adequately explain our data and
allow us to characterize multiple fracture sets.

\section{Theoretical background}

\subsection{Effective anisotropy induced by vertical cracks}

The effective compliance tensor, $\bm{s}_e$, of a fractured isotropic rock is generally
represented as the sum,
\begin{equation}\label{ru.eq.01}
  \bm{s}_e = \bm{s}_b + \Delta \bm{s} \, ,
\end{equation}
of the compliance, $\bm{s}_b$, of the unfractured background and the cumulative contribution,
$\Delta \bm{s}$, of multiple fractures. In the non-interaction approximation\cite{Kachanov1980, Kachanov1993, GrechkaKachanovORTfrac2006}, the nonzero elements of tensor $\Delta \bm{s}$ (in Voigt
notation) are
\begin{eqnarray}\label{ru.eq.02}
  \Delta s_{11} \Bs = \frac{16 \, {e}_1 \, (1 - \nu_b^2)}{3 \, E_b} \,
           (1 - \varsigma) \, , \quad
  \Delta s_{55} = \frac{32 \, {e}_1 \, (1 - \nu_b^2)}{3 \, E_b \, (2 - \nu_b)} \, ,
  \nonumber \\
  \Delta s_{22} \Bs = \frac{16 \, {e}_2 \, (1 - \nu_b^2)}{3 \, E_b} \,
           (1 - \varsigma) \, , \quad
  \Delta s_{44} = \frac{32 \, {e}_2 \, (1 - \nu_b^2)}{3 \, E_b \, (2 - \nu_b)} \, ,
  \\
  \Delta s_{66} \Bs  = \Delta s_{44} + \Delta s_{55} \, .
  \nonumber 
\end{eqnarray}
Here ${e}_1$ and ${e}_2$ are the densities of the two principal vertical fracture sets, $0
\leq \varsigma \leq 1$ is the fluid factor ($\varsigma \approx 0$ for dry cracks,
$\varsigma \approx 1$ for liquid-filled ones), and $E_b$ and $\nu_b$ are the Young's
modulus and Poisson's ratio of the background, respectively. They are given by the standard
expressions,
\begin{equation}\label{ru.eq.03}
  E_b = \mu_b \, \frac{3 \, \lambda_b + 2\, \mu_b}{\lambda_b + \mu_b} ~~ \text{and} ~~
  \nu_b = \frac{\lambda_b}{2 \, (\lambda_b + \mu_b)} \, ,
\end{equation}
in terms of the Lam\'{e} parameters $\lambda_b$ and $\mu_b$ of the host rock. The
compliance components in equations~\ref{ru.eq.02} are written in the coordinate frame
whose ${x}_1$ and ${x}_2$ axes are normal to the two principal fracture sets, and the
${x}_3$ axis is vertical.

The predictions of the non-interaction effective media theory (equations~\ref{ru.eq.01}
and~\ref{ru.eq.02}) appear to be remarkably simple. Yet, their high accuracy has been confirmed
numerically for fracture arrays that grossly violate the assumptions of the non-interaction
approximation\cite{GrechkaKachanovORTfrac2006, GrechkaKachanovXfrac2006, GrechkaetalSfrac2006}. Specifically,
formulations~\ref{ru.eq.01} and~\ref{ru.eq.02} remain sufficiently accurate for multiple sets of
differently oriented, irregularly shaped, partially closed cracks that might intersect each other
and form interconnected fracture networks. This means that long (compared to the crack sizes)
seismic waves propagating through those fractures ``see'' them as if they were isolated and
orthogonal to each other. As a result, the effective symmetry sensed by such long seismic waves is
nearly orthorhombic even though no local symmetry exists on the scale of a few fractures. This
conclusion leads to devising an efficient scheme for inverting the fracture parameters from seismic
data.

\subsection{Inversion strategy}

As follows from equations~\ref{ru.eq.01}~--~\ref{ru.eq.03}, the effective elastic properties
of rocks with multiple vertical fracture sets are described by the parameter vector
\begin{equation}\label{mfs.eq.50a}
  \bm{m} = \big\{\lambda_b, \, \mu_b, \, {e}_1, \, {e}_2, \, \varsigma \big\} \,
\end{equation}
that might vary spatially on the scale of seismic survey. Grechka and Kachanov\cite{GrechkaKachanovORTfrac2006} demonstrated that all components of $\bm{m}$ can be uniquely estimated from 3D, multiazimuth, multicomponent
seismic reflection data. (Of course, seismic data constrain the density-normalized background
Lam\'{e} constants rather than $\lambda_b$ and $\mu_b$ themselves.) In particular, it has been
shown that $\bm{m}$ can be unambiguously inverted from data
\begin{equation}\label{mfs.eq.51a}
  \bm{d}(\bm{m}) =
  \left\{ \frac{V_{S1}}{V_{P0}}, \, \frac{V_{S2}}{V_{P0}}, \,
  \bm{W}^{P}, \, \bm{W}^{S1}, \, \bm{W}^{S2}  \right\} ,
\end{equation}
where $V_{P0}$, $V_{S1}$, and $V_{S2}$ are the velocities of the vertically propagating P- and two
split shear-waves (the fast S$_1$ and the slow S$_2$). The velocity ratios entering $\bm{d}$ can be
computed from the zero-offset times after establishing the P-to-S event correspondence. The
$2 \times 2$ matrices $\bm{W}$ in equation~\ref{mfs.eq.51a} are the normal-moveout (NMO) ellipses
of pure modes reflected from a horizontal interface\cite{GrechkaTsvankin1998}. 

Vertical cracks result in effective orthorhombic media with two vertical symmetry planes,
therefore, all three NMO ellipses $\bm{W}^{P}$, $\bm{W}^{S1}$, and $\bm{W}^{S2}$ are
always co-oriented for a horizontal reflector, their axes pointing along the normals to the two
principal fracture sets. Consequently, data vector $\bm{d}$ given by equation~\ref{mfs.eq.51a} contains 
eight quantities: two velocity ratios and three pairs of the ellipse semi-axes~| the
symmetry-direction NMO velocities of the P-, S$_1$-, and S$_2$-waves. Only seven quantities of those
eight are independent because two out of four semi-axes of the shear-wave NMO ellipses coincide\cite{GrechkaTheophanisTsvankin1999}. Thus, the functional dependence $\bm{d}(\bm{m})$ 
consists of seven equations and five unknowns. To aid understanding of this dependence, 
Appendix~\ref{AppA} gives the weak-anisotropy (or small-crack-density) approximations of both the 
elements of $\bm{d}$ and the anisotropy coefficients of orthorhombic media\cite{Tsvankin1997}. 

In practice, we compute data $\bm{d}$ in equation~\ref{mfs.eq.51a} from wide-azimuth P, S$_1$, and S$_2$
common-midpoint (CMP) gathers. The NMO ellipses are obtained from 3D semblance 
analysis\cite{GrechkaTsvankin1999, VasconcelosTsvankin2006} that maximizes the semblance by fitting an
ellipse to the azimuthal variation of the NMO velocity. Another technique used in the industry is
based on measuring the traveltime shifts of a given reflection event after azimuthally-independent
NMO correction\cite{Jenner2001PhD}. Once the azimuthal NMO analysis is completed, the Dix-type
differentiation\cite{GrechkaTsvankin1999} is applied to calculate the interval NMO ellipses
$\bm{W}_{\rm int}^{P}$, $\bm{W}_{\rm int}^{S1}$, and $\bm{W}_{\rm int}^{S2}$ that enter
the interval data vector $\bm{d}_{\rm \, int}$. These ellipses typically exhibit some
misalignments due to a variety of factors such as noise, possible azimuthal anisotropy of the host
rock, reflector dip, fracture tilt, and interaction of the closely spaced cracks~| none of them
captured in our model. The data indicate, however, that those misalignments are usually small (see
Figure~\ref{ffmis} and the related discussion); therefore, we ignore them and estimate the
parameter vector $\bm{m}$ at each CMP location though nonlinear minimization,
\begin{equation}\label{misfit}
    \mathcal{F} = \underset{(\bm{m})}{\rm min} \,
        || \bm{d}_{\rm \, int} - \bm{d}(\bm{m}) || \, ,
\end{equation}
that operates with semi-axes of the ellipses only. In our inversion procedure, we use the exact
expressions for $\bm{d}(\bm{m})$ rather than their small-crack-density approximations given
in Appendix~\ref{AppA}. Finally, we compare the inferred crack orientations with the available
borehole measurements.

\section{Rulison field data}

\subsection{Brief overview of the field}

\begin{figure}
	\centerline{\begin{tabular}{c}
			\includegraphics[width=8.5cm]{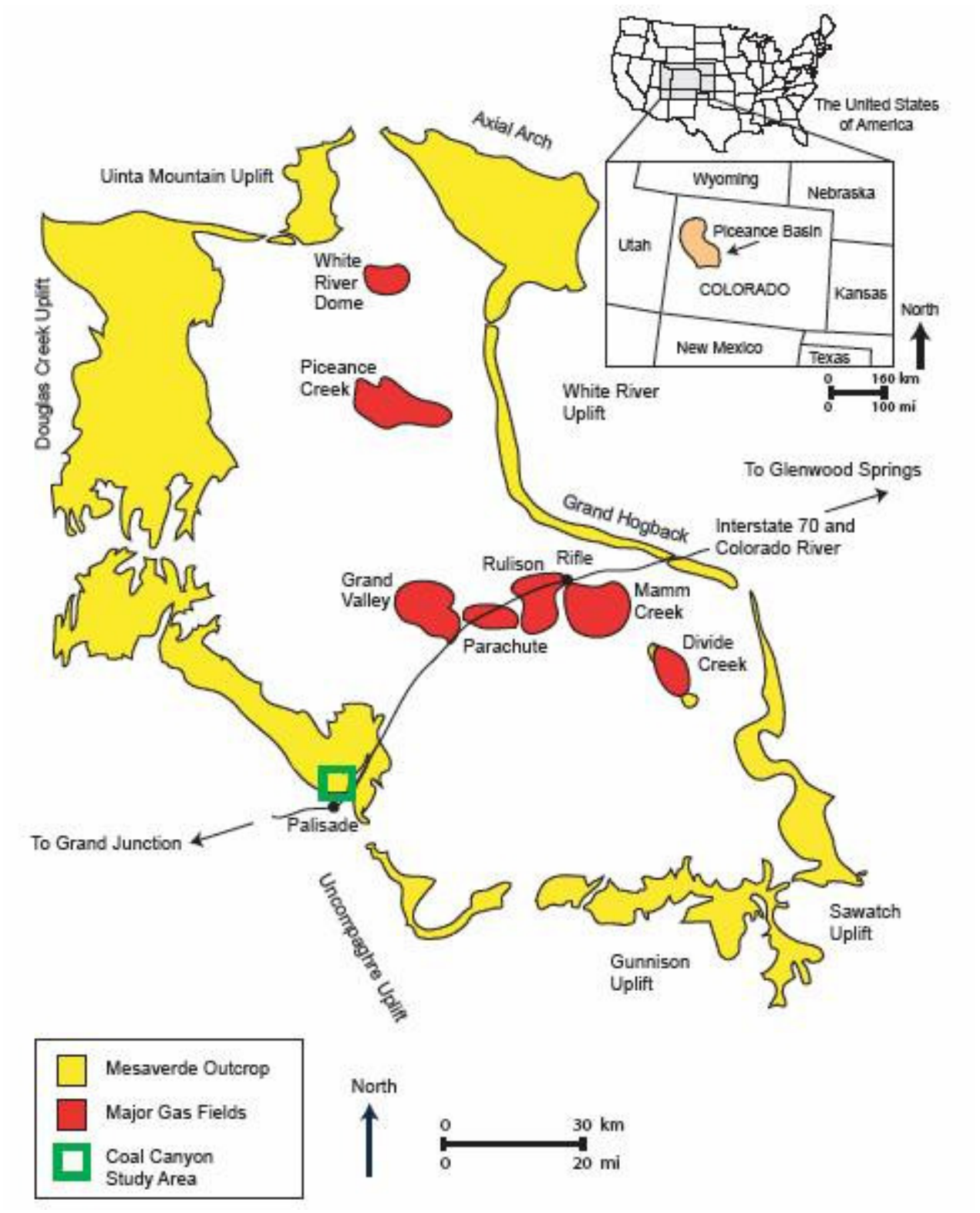} \\
	\end{tabular}}
	\caption{Map of gas-producing fields in Piceance Basin, Colorado (after Ellison\cite{Ellison2004}). }
	\vspace{0.0cm} \label{map}
\end{figure}

\begin{figure}[ht]
\centerline{\begin{tabular}{c}
  \includegraphics[width=10.0cm]{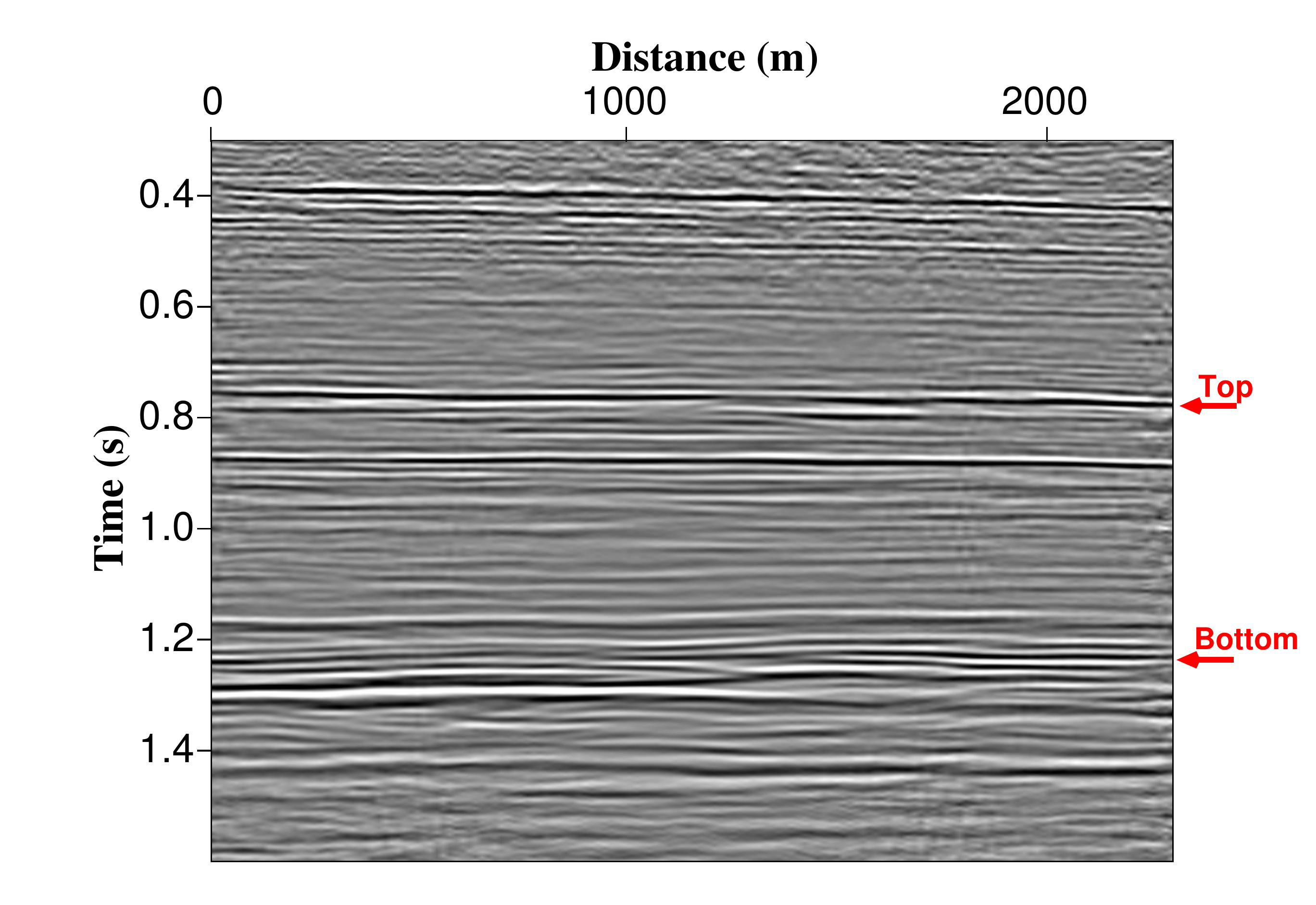} \\
\end{tabular}}
\caption{P-wave seismic section in east-west direction. Arrows mark the reflection events
used for azimuthal velocity analysis. } 
\vspace{0.0cm} \label{section}
\end{figure}

The Rulison Field, located in the South Piceance Basin, Garfield county, Colorado (Figure~\ref{map}),
produces gas from a 400~m thick pay section of channel sand lenses, stacked within the Cretaceous
sandstones of the Mesaverde group. Porosities (from 1\% to 10\%) and permeabilities (from 1~$\mu$D
to 60~$\mu$D) of these sand bodies characterize the formation as a tight gas reservoir\cite{Rojas2005}. Because of such a low matrix permeability, it is believed\cite{Johnsonetal1987, Spencer1989, Lynnetal1999} that gas production is controlled by open natural fractures. Previous studies\cite{Lorenz2004, Jansen2005} suggest that fracture orientations vary over the field, and their intersections are thought to be important for the reservoir production. The average thickness of sand lenses is about 3~m, making them impossible to image with surface seismic data. As
neither sand bodies nor fracture zones can be individually isolated on seismic images, it is
necessary to characterize the fractures collectively as an effective medium. Reflector dips at the
Rulison are small and can be neglected for the purposes of seismic data processing
(Figure~\ref{section}).

\subsection{Data acquisition}

A 3D, multicomponent seismic data set covering 2.2~km $\times$ 2.5~km was acquired by the Reservoir
Characterization Project (Colorado School of Mines) in October, 2003 to characterize the Rulison Field
with information from multiple sources (surface seismic, vertical seismic profiling, geomechanics, well logs, production data) and to monitor the changes in production and stresses with time-lapse measurements. For our
study, we use the baseline wide-azimuth, 9C seismic survey. The sources are 3C (vertical and
horizontal) vibrators placed at 33.5~m (110~ft) intervals along the lines oriented east-west; the
crossline spacing is approximately 201~m (660~ft). The 3C receivers, 1500 Vectorseis$^\circledR$
digital accelerometers providing a total of 4500 live channels, are laid out in the north-south direction
with 33.5~m (110~ft) inline and 101~m (330~ft) crossline intervals. This acquisition design yields
an uncommonly high fold of about 500 in the middle of the survey and 200 to 300 on average.

\begin{figure}[h]
	\centerline{\begin{tabular}{c}
			\includegraphics[width=13.5cm]{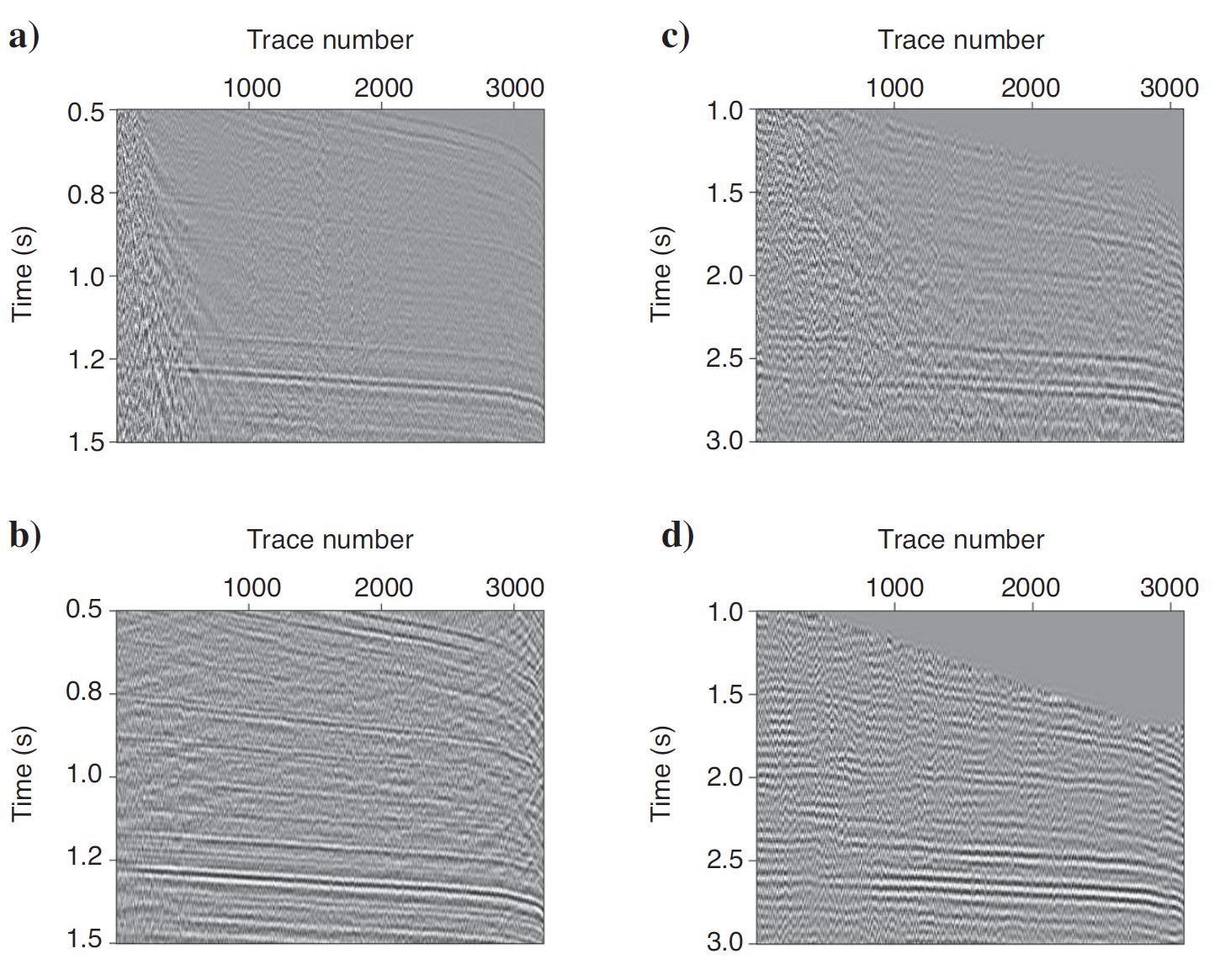} \\
	\end{tabular}}
	\caption{Representative input (a) and (c) and $f$-$k_x$-$k_y$-filtered (b) and (d) CMP gathers
		of the P- (left column) and S$_1$-waves (right column). Offsets increase with the increasing trace number. Both offsets and azimuths are irregularly sampled in the CMP domain.} \vspace{0.0cm} 
	\label{fk}
\end{figure}

\subsection{Data processing}

Our input data consist of three CMP-sorted volumes (P, S$_1$ and S$_2$), deconvolved for the source
signature and corrected for the source-receiver statics. These pre-processing steps have been carried out
by Veritas DGC prior to our project. In addition, ground-roll suppression turned out to be necessary.
Strong surface waves, present in all three data volumes, are particularly harmful for the shear
reflections (Figure~\ref{fk}). As the S-wave velocities are slower than the P-wave ones, shear
events come later in time, and their greater offset interval is covered by the ground-roll (compare
Figures~\ref{fk}a and~\ref{fk}c). To suppress it, we resort our irregularly sampled (in offsets and
azimuths) data into shot and receiver lines, where the sampling is regular, and apply an
\mbox{$f$-$k_x$-$k_y$} filter that preserves the slopes up to 0.5~s/km. This approach to
ground-roll suppression does not destroy azimuthal variations of traveltimes and results in a
significant improvement of the reflected events (Figure~\ref{fk}).

Another pertinent issue concerns the two shear-wave volumes (S$_1$ and S$_2$) that we have at our
disposal. The shear data have been rotated to the azimuth of the principal regional stress at N45E.
Treating this azimuth as the orientation of one of the principal fracture sets implies a constant
fracture azimuth throughout the field, in contradiction with the laterally variable orientations of
the P-wave NMO ellipses observed from the bottom of the producing interval. To mitigate this
discrepancy, we rotate the shear volumes to the principal directions of these P-wave NMO ellipses
at each CMP location prior to performing the S-wave velocity analysis. The weak P-wave azimuthal
anisotropy in the overburden, quantified by the eccentricities of the P-wave NMO ellipses consistently smaller
than 3\%, implies that fractures in the reservoir are the main source of the observed azimuthal
anisotropy. If the latter is consistent with our theoretical model, the performed rotation brings
the shear-waves to their principal directions; if not, such a rotation introduces some errors.
These errors, however, are small for the majority of our data, as demonstrated in Figure~\ref{ffmis} below.

After the ground-roll suppression and shear-wave rotation, we resort the data back to CMP
geometry, create $9 \times 9$ (135~m$\, \times \,$135~m) superbins and extract the NMO
ellipses over the survey area. At each superbin location, estimation of the NMO ellipses is
carried out around the zero-offset times that correspond to the reservoir top and bottom
horizons (Figure~\ref{section}), interpreted from the P, S$_1$, and S$_2$ image volumes.
The NMO ellipses are computed at 5~ms time increments over a window centered at the
horizon times. The averaging lengths of the semblance operators are 35 ms for the P-wave and 105
ms for the shear-wave data. Our use of the interpreted horizons for velocity analysis ensures the consistency of results with the geologically-based time structure.

\begin{figure}
\centerline{\begin{tabular}{c}
  \includegraphics[width=15.0cm]{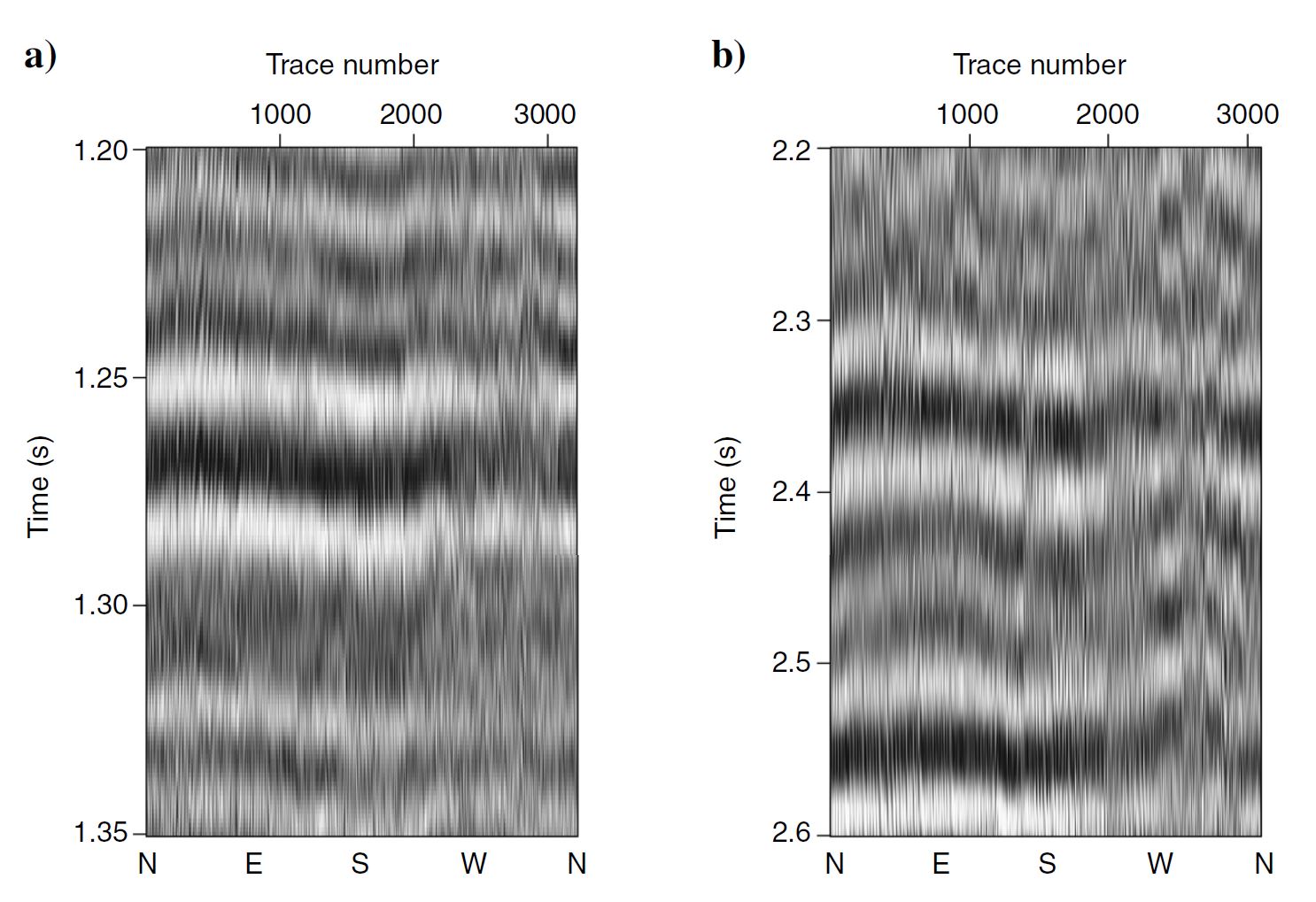} \\
\end{tabular}}
\caption{P- (a) and S$_1$-wave (b) gathers after the azimuthally-invariant NMO correction. The
traces are sorted by the source-receiver azimuth. The apparent cosine-type dependence of the
residual moveout is indicative of azimuthal anisotropy. The reflection events at approximately 
1.27~s in (a) and 2.35~s in (b) correspond to the reservoir bottom. } \vspace{0.0cm}
\label{resMO}
\end{figure}

We obtain the P-wave NMO ellipses over the entire survey area, the corresponding
semblance values ranging from 0.55 to 0.65. As the shear data are noisier (compare
Figure~\ref{fk}b with~\ref{fk}d), the S-wave NMO ellipses have been estimated for only
about one third of the area. We discard the S-wave NMO ellipses whenever the
semblances fall below~0.25. Indeed, the reflection events cannot be
visually distinguished on the gathers with such low semblance values. On the other hand,
in the higher-quality data areas, the azimuthal traveltime variations are clearly seen on
the gathers corrected with an azimuthally-constant NMO velocity (Figure~\ref{resMO}).

After estimating the P, S$_1$, and S$_2$ NMO ellipses for the top and bottom horizons over the
entire survey area, we proceed with the Dix-type differentiation\cite{GrechkaTsvankin1999} that 
yields the interval ellipses $\bm{W}_{\rm int}^{P}$, $\bm{W}_{\rm int}^{S1}$, and
$\bm{W}_{\rm int}^{S2}$. Finally, minimizing the objective function in equation~\ref{misfit}, we
invert $\bm{W}_{\rm int}^{P}$, $\bm{W}_{\rm int}^{S1}$, and $\bm{W}_{\rm int}^{S2}$ and
the interval zero-offset time ratios for the fracture parameters.

\subsection{Fracture characterization}

\begin{figure}
\centerline{\begin{tabular}{c}
  \includegraphics[width=16.0cm]{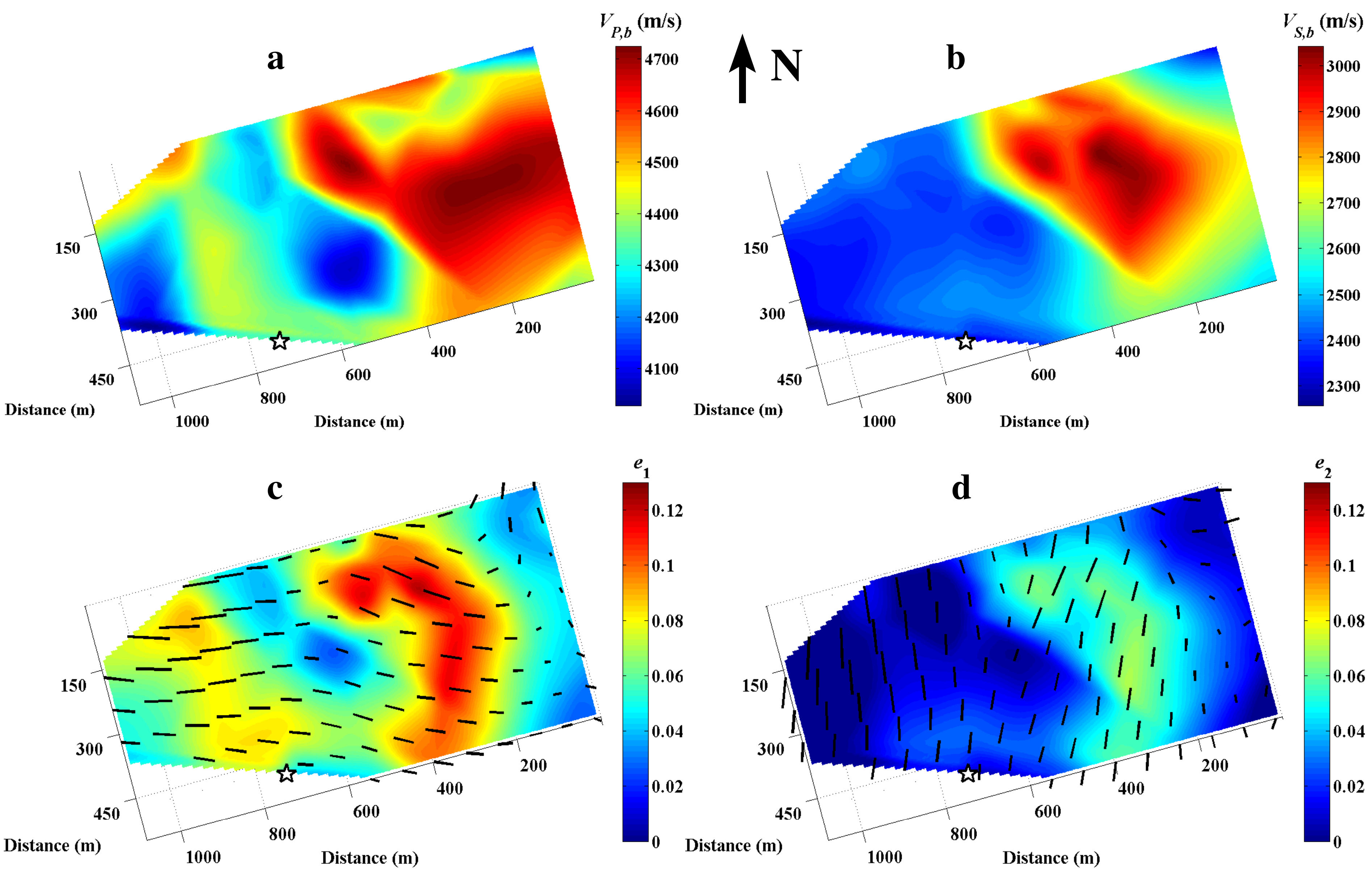} \\
\end{tabular}}
\caption{Output of the fracture characterization: the background velocities $V_{P,\,b}$ and
$V_{S,\,b}$ of the P- (a) and S-waves (b), and the principal crack densities $e_1$~(c) and $e_2$~(d).
The directions of the principal fracture sets are shown with the short black lines, their lengths proportional
to the eccentricities of the interval P-wave NMO ellipses (shown in Figure~\ref{rul.fig.04}b).
These maps (as well as those in Figures~\ref{rul.fig.02}--\ref{rul.fig.04}) are the result of
bi-cubic interpolation of the original data points spaced on a $9 \times 9$ (135~m$\, \times
\,$135~m) superbin grid. The stars indicate the location of well B.\ RWF 524-20, where an FMI log has been
acquired (Figure~\ref{frac}).} \vspace{0.0cm} \label{medpar}
\end{figure}

Figure~\ref{medpar} displays the inverted background velocity fields $V_{P,\,b}$ and
$V_{S,\,b}$ and the principal crack densities
${e}_1$ and ${e}_2$ (by definition, ${e}_1 \geq {e}_2$). The crack densities in
Figure~\ref{medpar}c are considerably greater than those in Figure~\ref{medpar}d, 
suggesting that the fracturing is dominated by the cracks trending in the WNW-ESE
direction. Our results indicate that the western part of the area is mostly controlled by
a single fracture set that has the density ${e}_1$, whereas the eastern part has a
non-negligible contribution of other, differently oriented fractures that exhibit
themselves as the set with the crack density ${e}_2$ (Figures~\ref{medpar}c
and~\ref{medpar}d). We also observe that the areas of the highest crack densities do not
necessarily coincide with the largest P-wave NMO ellipse eccentricities and revisit this issue later.

The background velocities (Figures~\ref{medpar}a and~\ref{medpar}b) show a sizable contrast
between the eastern and western portions of the study area, with both P- and S-wave velocities
being slower in the west. The fluid factors $\varsigma$ have been estimated also but we do not
display them because all obtained values of $\varsigma$ are smaller than~0.01. These estimates
are consistent with the fact that the Rulison Field produces dry gas.

\subsection{Orthorhombic model of the reservoir}

\begin{figure}
\centerline{\begin{tabular}{c}
  \includegraphics[width=16.0cm]{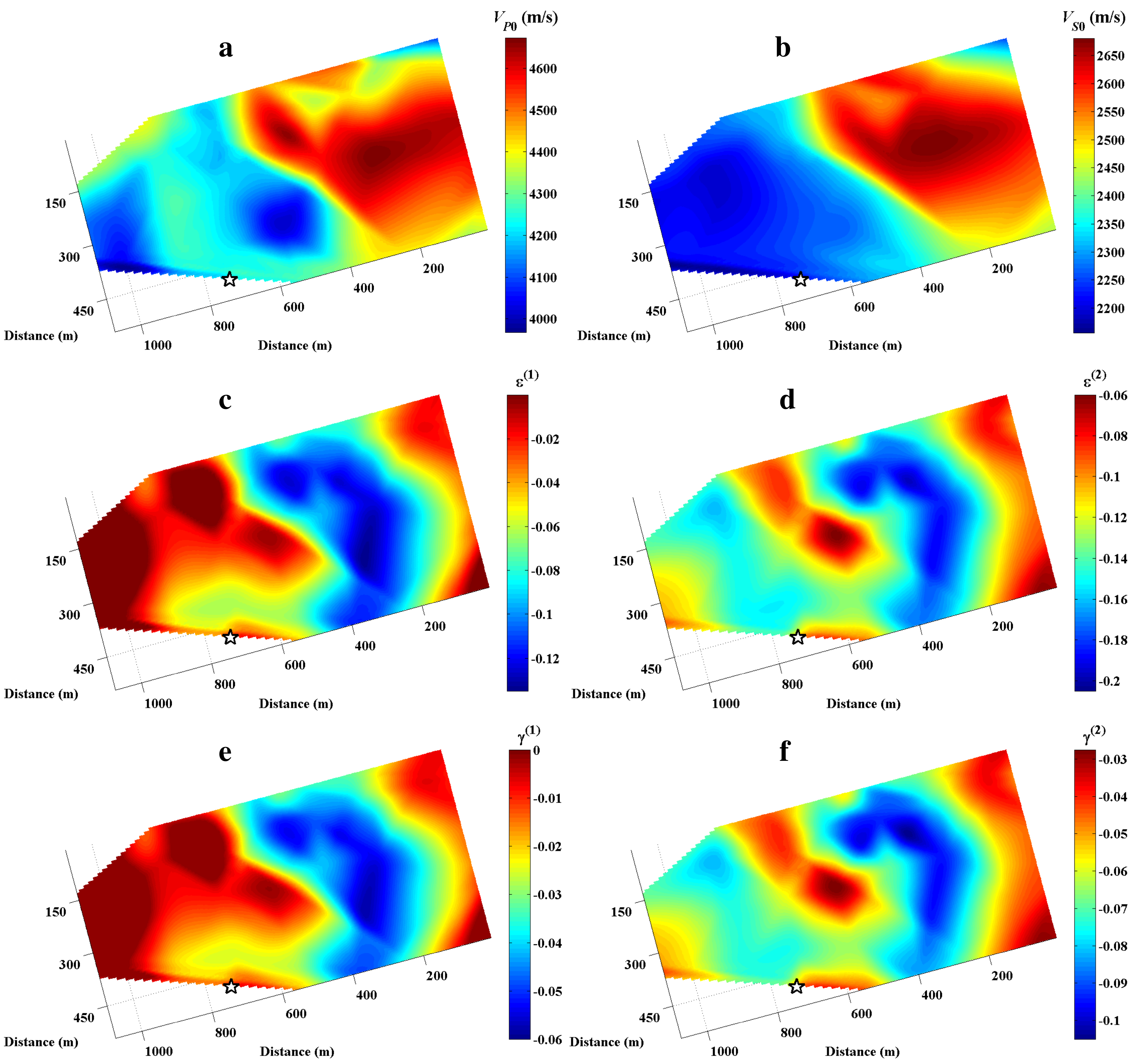} \\
\end{tabular}}
\caption{Vertical velocities (a) $V_{P0}$, (b) $V_{S0}$ and anisotropy coefficients (c)
$\epsilon^{(1)}$, (d)~$\epsilon^{(2)}$, (e) $\gamma^{(1)}$, and (f) $\gamma^{(2)}$ at
Rulison reservoir.} \vspace{0.0cm} \label{rul.fig.02}
\end{figure}

\begin{figure}
\centerline{\begin{tabular}{c}
  \includegraphics[width=16.0cm]{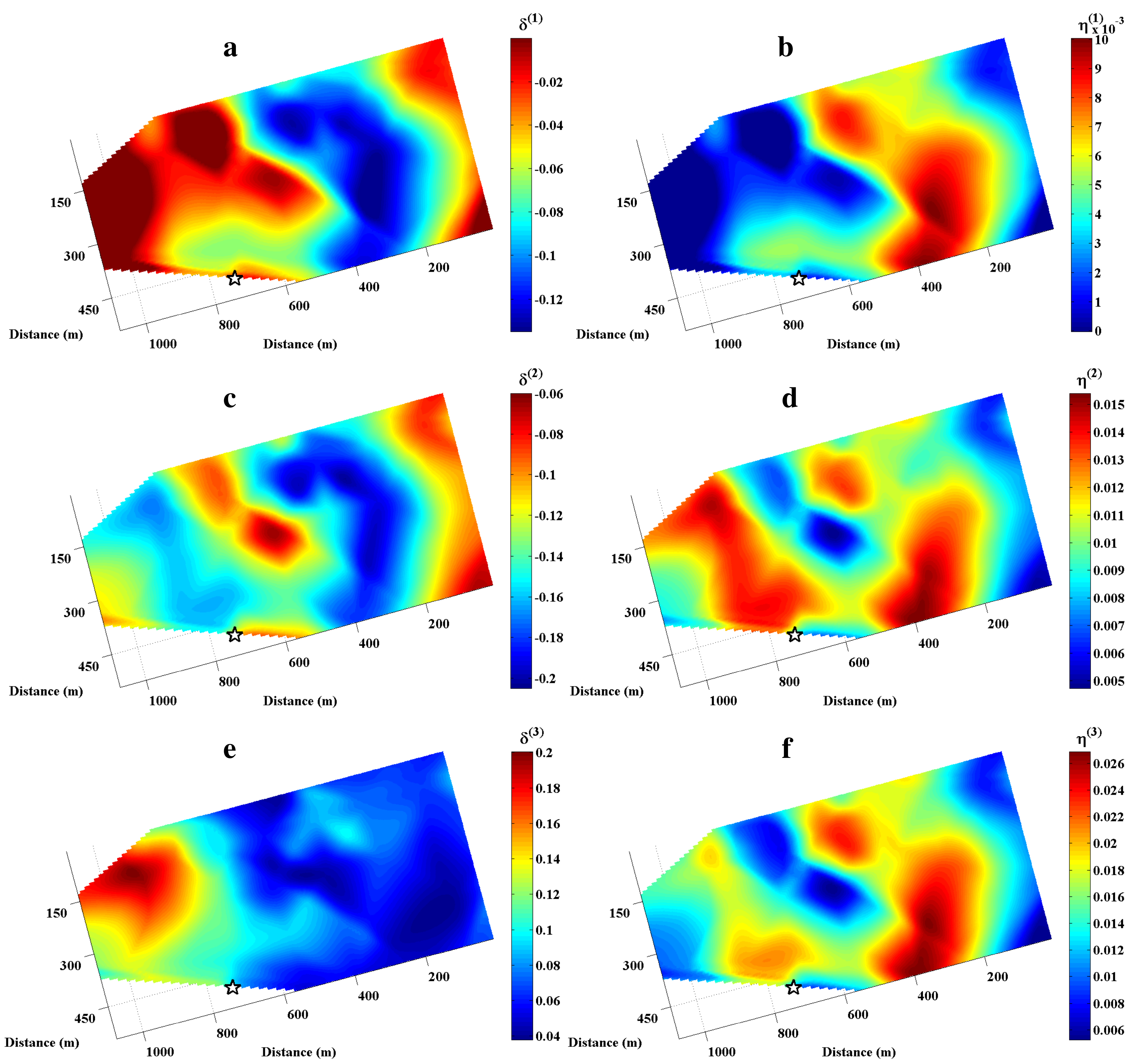} \\
\end{tabular}}
\caption{Anisotropy coefficients (a, c, e) $\delta^{(1,\,2,\,3)}$ and (b, d, e)
$\eta^{(1,\,2,\,3)}$ at Rulison reservoir.} \vspace{0.0cm} \label{rul.fig.03}
\end{figure}

The estimated background velocities and crack densities (along with the fluid factors
$\varsigma \approx 0$) are sufficient for building an {orthorhombic} depth model of
the reservoir. Figures~\ref{rul.fig.02} and~\ref{rul.fig.03} display the vertical velocities
$V_{P0}$ and $V_{S0}$ and Tsvankin's\cite{Tsvankin1997} anisotropy coefficients, their
small-crack-density approximations found in Appendix~\ref{AppA}. We also present the
anellipticity coefficients $\eta^{(1,\,2,\,3)}$ (the right column in Figure~\ref{rul.fig.03})
to demonstrate that the predicted anisotropy is nearly elliptical.

It is instructive to point out that the model in Figures~\ref{rul.fig.02} and~\ref{rul.fig.03} was
obtained from reflection data only without the use of any borehole information to constrain the vertical
depth scale. Although well-log or checkshot data are usually required for building orthorhombic
subsurface models in depth domain\cite{GrechkaPechTsvankin2005}, we did not need such data in our study. 
The reason is that we explicitly target the {crack-induced} rather than general orthotropy.
The former is significantly simpler because it is governed by fewer (five rather than nine)
independent parameters, the reduction in the number of unknowns allowing us to rely solely on
surface seismic data for building the unique orthorhombic model of the Rulison reservoir. As
Figures~\ref{rul.fig.02} and~\ref{rul.fig.03} indicate, the reservoir is noticeably anisotropic:
the magnitudes of anisotropy coefficients $\epsilon^{(2)}$ and $\delta^{(2)}$ reach~0.2 at the
highest total crack density of $e = e_1 + e_2 = 0.17$ (Figures~\ref{medpar}c and~\ref{medpar}d).

\subsection{Conventional approaches to fracture characterization}

\begin{figure}[t]
\centerline{\begin{tabular}{c}
  \includegraphics[width=16.0cm]{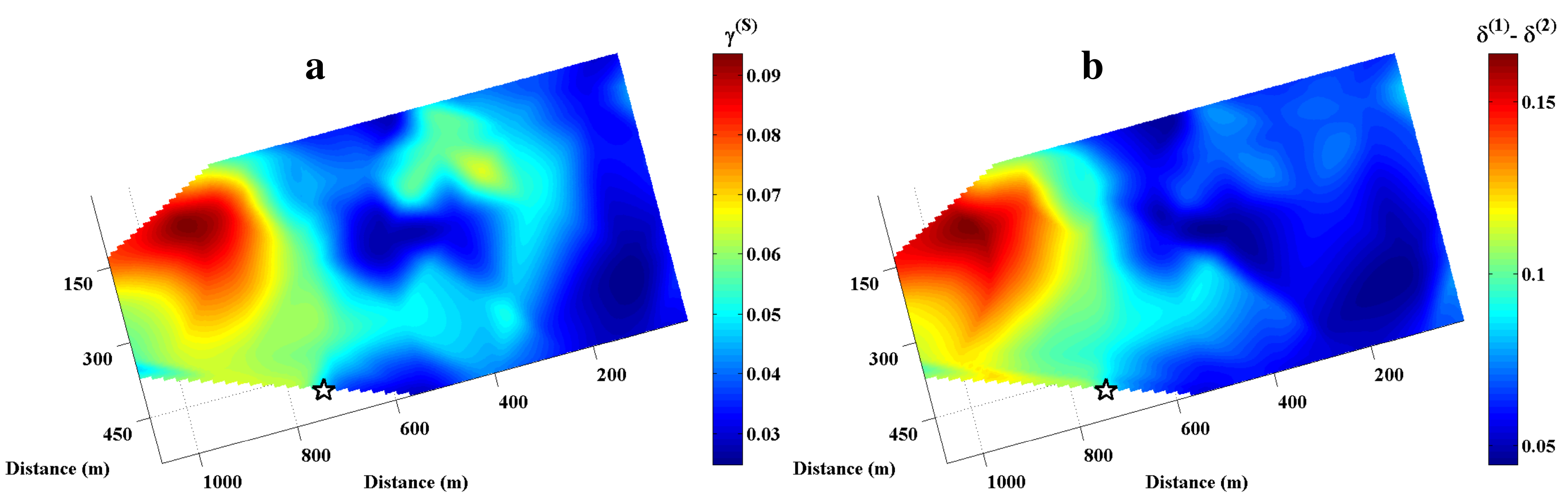} \\
\end{tabular}}
\caption{The shear-wave splitting coefficient $\gamma^{(\rm S)}$ (a) and the eccentricity
of the P-wave NMO ellipses $\delta^{(1)} - \delta^{(2)}$ (b) at Rulison reservoir. }
\vspace{0.0cm} \label{rul.fig.04}
\end{figure}

Having estimated the reservoir parameters from both P- and S-wave data, we can predict
what would happen if we relied on either S- or P-waves alone for fracture
characterization. In a typical pure shear-wave survey, one would measure the shear-wave
splitting coefficient, \mbox{$\gamma^{(\rm S)} \approx \gamma^{(1)} - \gamma^{(2)}$,} and
interpret it as the crack density of a single fracture set. The result of this
interpretation in Figure~\ref{rul.fig.04}a suggests that the western part of the study area
is more fractured than the eastern one, quite opposite to the conclusion drawn from
Figures~\ref{medpar}c and~\ref{medpar}d.

If only the P-wave data were used\cite{GrechkaTsvankin1999}, one would estimate the
eccentricity of the P-wave NMO ellipses, quantified by the difference of two
$\delta$ coefficients, \mbox{$\delta^{(1)} - \delta^{(2)}$.} This difference, shown in
Figure~\ref{rul.fig.04}b, yields a similar result that the western part of the area is
apparently more fractured than the eastern part.

The origin of these mutually contradictory conclusions can be understood from 
approximations~\ref{ddel} and~\ref{gams}. Clearly, both quantities $\gamma^{(\rm S)}$ and
$\delta^{(1)} - \delta^{(2)}$ are proportional to the difference $\, e_1 - e_2 \,$ between the
crack densities of two principal fracture sets. As a consequence, the shear-wave splitting
coefficients and the eccentricities of the P-wave NMO ellipses are useful for fracture
characterization only if one fracture set dominates, that is, when $e_1 \gg e_2$. If multiple sets
of cracks resulting in comparable $e_1$ and $e_2$ are present in the subsurface, both differences
$\gamma^{(1)} - \gamma^{(2)}$ and $\delta^{(1)} - \delta^{(2)}$ become ambiguous. In particular, if
the two principal crack densities coincide, $e_1 = e_2 \neq 0$, both differences vanish,
$\gamma^{(1)} - \gamma^{(2)} = \delta^{(1)} - \delta^{(2)} = 0$, and one would arrive to an
obviously erroneous conclusion of the absence of fractures.

\section{Verification of the results}

An important part of any inversion procedure is assessment of errors in the estimated quantities.
This section describes our efforts in this direction.

\subsection{Variances in data and estimated parameters}

We estimated the data variances associated with uncertainties in traveltime picking. A plausible
picking interval was defined in accordance with the auto\-correlation function of traces in a CMP
gather. Specifically, we take this interval to be equal to the time lag at which the autocorrelation
drops by a factor of~10 from its zero-lag value. Such a definition yields the picking uncertainties
equal to~8~ms for P- and 14~ms for S-waves. Next, a standard linear error-propagation technique is
applied to translate these traveltime errors into uncertainties in the NMO ellipses. Finally, using
the Fr\'{e}chet-derivative matrix $\partial d_j / \partial m_i$, where $m_i$ and $d_j$ are the
components of $\bm{m}$ and $\bm{d}$ (equations~\ref{mfs.eq.50a} and~\ref{mfs.eq.51a}), we
propagate the errors into the estimated quantities shown in Figure~\ref{medpar}. This yields
uncertainties of about 7\% in the velocities $V_{P,\,b}$ and $V_{S,\,b}$ and 0.01 in the principal
crack densities $e_1$ and $e_2$ at a 90\% confidence interval. Given that eccentricities of
the interval NMO ellipses are not large, we find the average errors in the ellipse azimuths to be equal to
$12^{\circ}$, $17^{\circ}$, and $22^{\circ}$ for the P-, S$_1$-, and S$_2$-waves, respectively.

Estimation of the model variances is not the only resource for validating our results. As
we have already mentioned, the fluid factors $\varsigma$ turned out to be nearly zero at
all CMP locations, as expected because of the absence of production of liquids (oil
or water) at the Rulison.

\begin{figure}[t]
\centerline{\begin{tabular}{cc}
  \includegraphics[width=7.0cm]{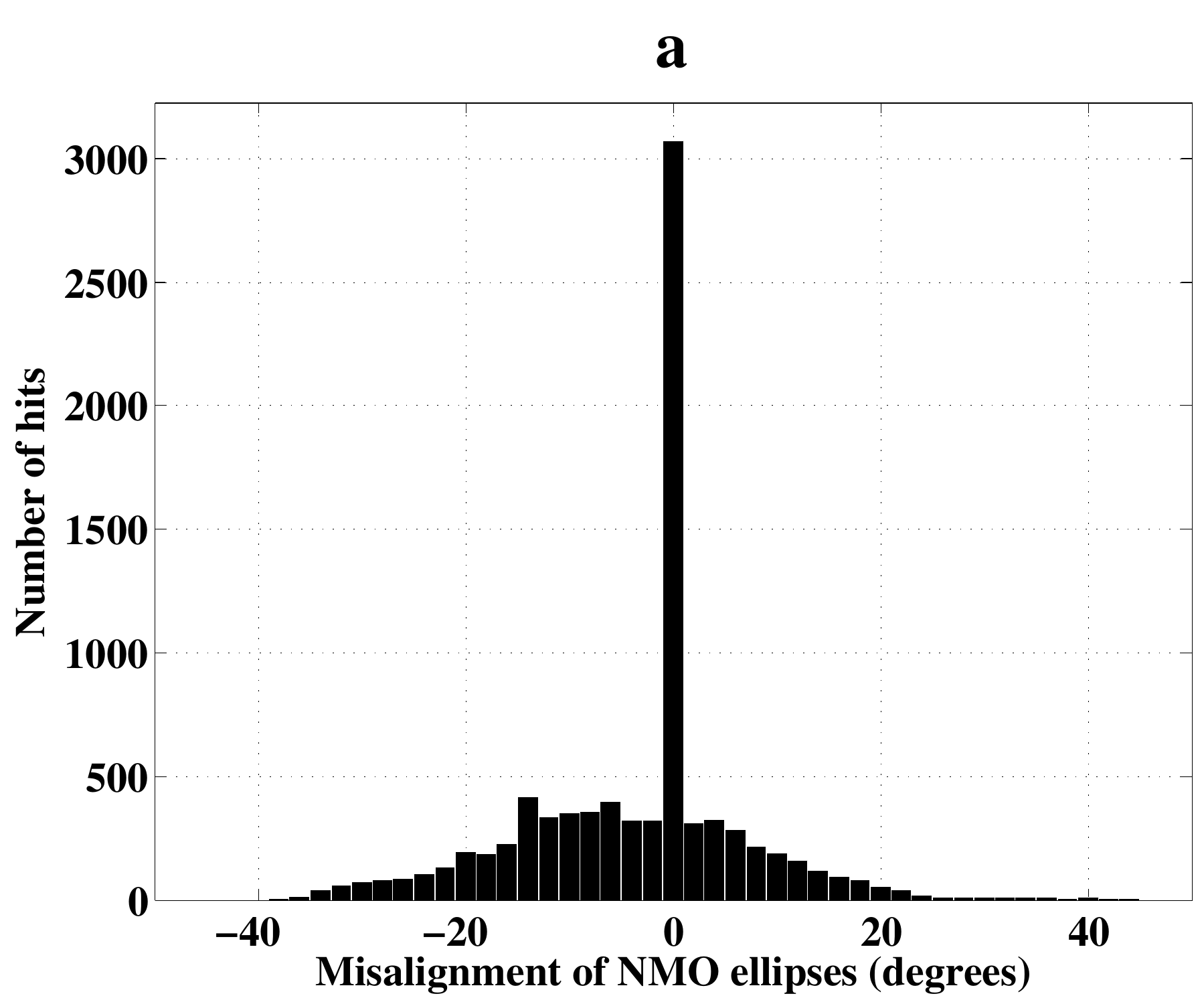} &
  \includegraphics[width=7.0cm]{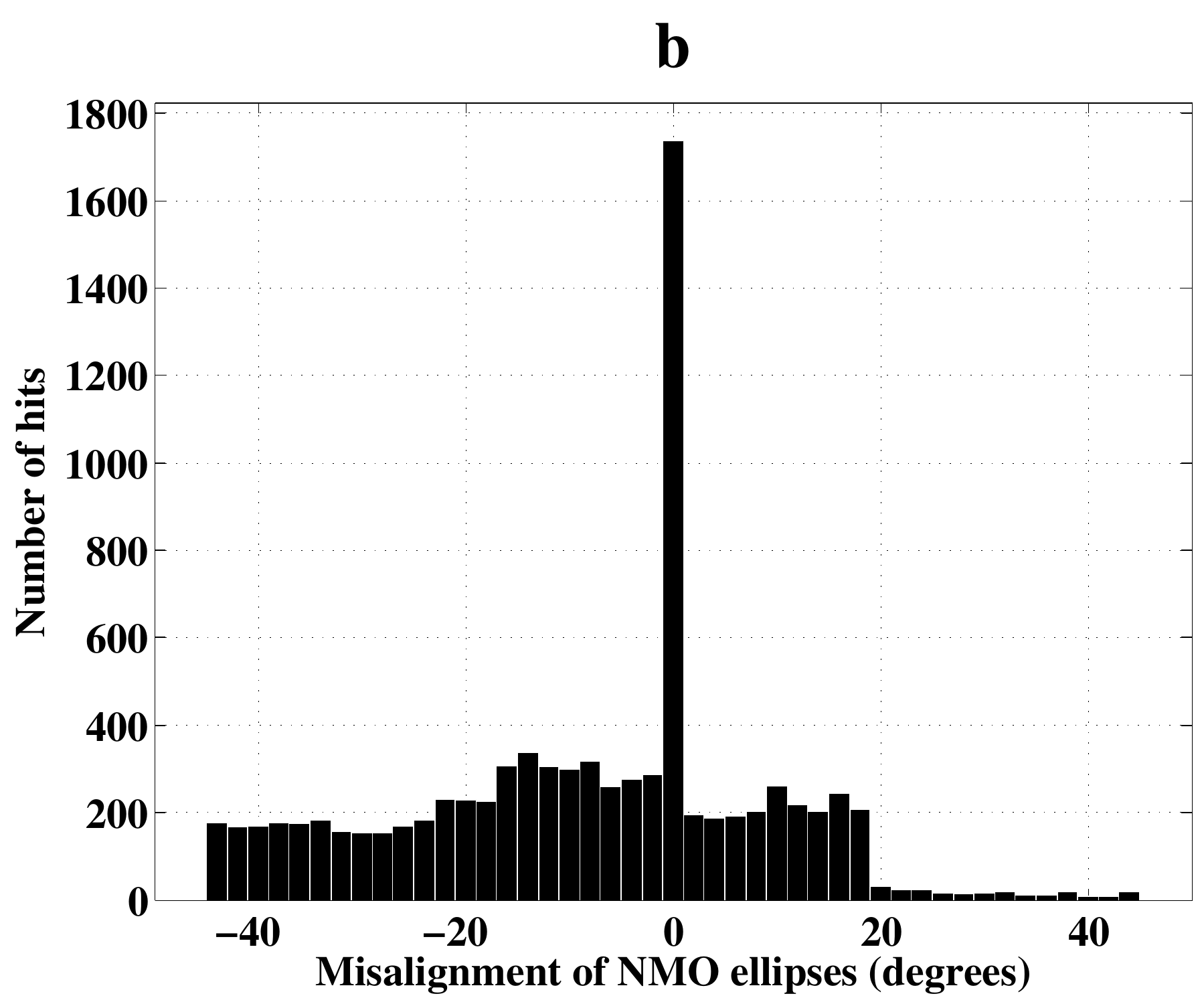} \\
\end{tabular}}
\caption{Histograms of azimuthal misalignments of the interval NMO ellipses of P- and S$_1$-waves
(a) and P- and S$_2$-waves (b). The numbers along the vertical axes correspond to the interpolated
values used to display the fracture-characterization results in
Figures~\ref{medpar}~--~\ref{rul.fig.03}.} \vspace{0.0cm} \label{ffmis}
\end{figure}

Another way of assessing the quality of our model assumptions is to analyze the misalignments in
the orientations of the interval NMO ellipses for different wave modes. The histograms in
Figure~\ref{ffmis} exhibit pronounced peaks at zero, indicating that the ellipses are aligned
at many CMP locations. Even though the presence of outliers (likely caused by noise in the data)
makes average differences in the ellipse orientations of the P- and S$_1$-waves and the P- and
S$_2$-waves be equal to~12$^\circ$ and~17$^\circ$, respectively, significant portion of the differences can be
attributed to fracture interaction or, in seismic terminology, to multiple scattering of elastic
energy by closely spaced cracks. For instance, Grechka and Kachanov\cite{GrechkaKachanovORTfrac2006} found this
mechanism to be responsible for the NMO ellipse misalignments up to 12$^\circ$ at the total crack
density $e = 0.14$. Thus, we conclude that orientations of the NMO ellipses are consistent with our
fracture model over most of the data.

\begin{figure}
	\centerline{\begin{tabular}{c}
			\includegraphics[width=8.0cm]{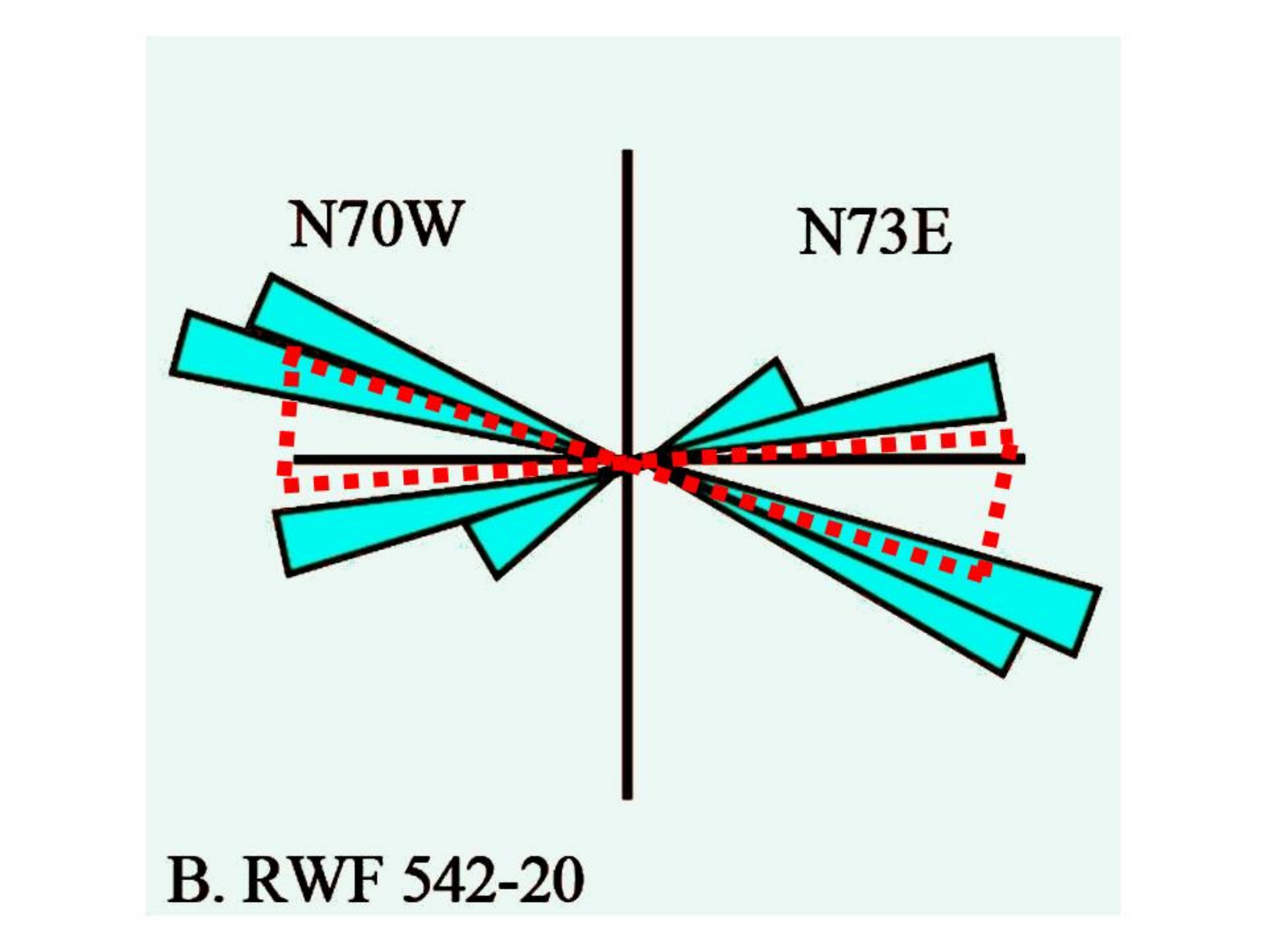} \\
	\end{tabular}}
	\caption{Fracture count (cyan) in well \mbox{B.\ RWF 542-20} (marked by the star in
		Figures~\ref{medpar}~--~\ref{rul.fig.04}) and the 90\% confidence interval (dotted red)
		corresponding to the azimuth of the fracture set with density ${e}_1$ estimated from
		seismic data.} \vspace{0.0cm} \label{frac}
\end{figure}

\subsection{Borehole measurements}

The survey area contains well \mbox{B.\ RWF 542-20} (the star in
Figures~\ref{medpar}~--~\ref{rul.fig.04}), where an FMI log was acquired, and fractures in the entire
reservoir were counted. Figure~\ref{frac} compares the borehole fracture count with our estimate of
the crack orientation. As a dominant set of cracks has been identified at the well location
(Figures~\ref{medpar}c and~\ref{medpar}d), we do not display the second set in Figure~\ref{frac}, 
for it has nearly zero crack density. The fractures observed in the FMI form two sets oriented
approximately at N70W and N73E (cyan in Figure~\ref{frac}). Clearly, these sets are not orthogonal
to each other; yet, their influence on the propagation of long (compared to the fracture sizes)
seismic waves is equivalent to that of two orthogonal sets. The dotted red line in
Figure~\ref{frac} shows the azimuth of the dominant equivalent set estimated from seismic.

Figure~\ref{frac} can be regarded as an illustration of the resolution achievable from
seismic data: while long seismic waves cannot resolve each individual fracture set, they
are sensitive to all fractures simultaneously. This cumulative influence of all cracks
on elastic wave propagation is described by the crack-density tensor\cite{Kachanov1980, GrechkaKachanovORTfrac2006} that can be represented in terms of
contributions of mutually orthogonal (or principal) fracture sets that control the seismic signatures and can be estimated from them. Thus, returning to Figure~\ref{frac}, we state that our fracture-characterization
results are consistent with the FMI log in well \mbox{B.\ RWF 542-20}.

\section{Discussion and conclusions}

We have discussed seismic characterization of fractures at the Rulison Field, Colorado. To
the best of our knowledge, Rulison is the first case study described in the geophysical
literature, where parameters of {multiple} fracture sets~| their
orientations, crack densities, and the type of infill~| have been estimated from
multicomponent seismic reflection data. The key in our parameter-estimation methodology
is the understanding, confirmed by extensive numerical modeling, that differently
oriented, multiple vertical cracks appear as two orthogonal sets for long seismic waves.
Following this understanding, we have targeted the parameters of those two orthogonal
(or principal) sets in our inversion.

The obtained results point to a single fracture set oriented in the WNW-ESE direction in
the western part of the study area and to two or more sets of cracks in its eastern part.
These multiple sets yield the principal crack densities reaching \mbox{$e_1 = 0.11$}
and $e_2 = 0.06$ and implying an interconnected fracture network. This conclusion follows
from a straightforward geometric consideration: it is extremely difficult to place two
sets of orthogonal fractures that have random locations and the crack densities of
$e_1 = 0.11$ and $e_2 = 0.06$ in a rock volume without letting those fractures
intersect each other. Thus, our inversion has identified places in the reservoir, where
swarms of intersecting cracks are likely to be present. Although multiple, differently
oriented fracture sets might be viewed as contradictory to a common belief that only a
single set of vertical open cracks directed along the regional stress field should exist
in the subsurface, multiple fracture sets misaligned with the present stress directions 
have been observed in the Piceance Basin\cite{Laubachetal1998, Lorenz2004} and 
other areas\cite{Laubachetal2004}.

As a by-product of our fracture-characterization procedure, we have built an
orthorhombic depth model of the Rulison reservoir. The reconstructed orthorhombic
velocity model explicitly attributes the observed anisotropy to the presence of
fractures. This rock-physics constraint reduces the number of independent quantities to
be estimated and obviously improves the stability of inversion. In particular, it allows
the vertical P- and S-wave velocities and all anisotropy coefficients to be uniquely
determined solely from 3D, multiazimuth, multicomponent seismic reflection data. We
believe that the Rulison is the first example of building an orthorhombic subsurface
velocity model from seismic reflection data without using borehole information.

We paid special attention to verifying the obtained results. Here we list the supporting evidence.

\begin{description} \vspace{-1mm}
    \item ~~(1) We propagated uncertainties in the picked zero-offset times and
    NMO ellipses into the quantities that characterize the fractures. This yielded
    errors of about~7\% in the P- and S-wave velocities in the host rock and around 0.01
    in the principal crack densities.

    \item ~(2) The Rulison Field produces dry gas. While knowing this beforehand, we intentionally
    included the fluid factor into our unknowns. Its estimated values came out to be
    nearly zero in the entire study area, consistent with our prior information about of the type of
    fracture infill.

    \item ~(3) The NMO ellipses of the P-, S$_1$-, and S$_2$-waves recorded from horizontal
    reflectors are supposed to be co-oriented according to the adopted non-interaction
    theory of crack-induced anisotropy. The ellipse misalignments observed in seismic
    data turned to be compatible with those predicted by numerical modeling that takes
    into account interaction of closely spaced fractures.

    \item ~(4) The available FMI log confirmed the principal fracture azimuth estimated from seismic.
\end{description} \vspace{-1mm}

Despite the above assurances that the obtained results not only exhibit internal
consistency but also fit some independent external data, comparison of our crack
densities with the estimated ultimate recoveries of wells in the study area shows little
correlation. The reason for that, we believe, is the physics of
fluid flow through fractured rocks. For low-porosity (1\% to 10\%) sandstones at the
Rulison, fracture networks appear to provide the main pathways for the gas flow. The
effective, crack-related permeability of these networks is essentially controlled by the
fracture {widths}, whereas the seismic signatures are governed by the crack
densities or the fracture {lengths}. If a relationship between the crack widths and
lengths is absent, the effective permeability and elasticity are uncorrelated. Perhaps, 
this is what we see at the Rulison.

Summarizing our seismic results, we reiterate that, to the best of our knowledge, our
paper is the first to demonstrate the following:

\vspace{-0.2cm}
\begin{itemize}
\item 3D, wide-azimuth, multicomponent seismic data can be used to characterize multiple
vertical fracture sets in a purely isotropic host rock. The estimated crack densities
might give an idea of where the fractures are likely to intersect and form
interconnected networks.

\item Orthorhombic subsurface models can be built from surface reflection data. We have
shown that for a special case of crack-induced orthotropy, allowing us to resolve the
time-depth ambiguity without relying on borehole data.
\end{itemize} \vspace{-0.2cm}

\section{Acknowledgments}

We thank the Reservoir Characterization Project (Colorado School of Mines) for providing seismic
data and Shell~E\,\&\,P for permission to publish the paper. We are grateful to Jon Sheiman (Shell)
for helping us with data processing, to Jorge Lopez (Shell), Ilya Tsvankin, Rodrigo Fuck (both
CSM), and anonymous reviewers for their comments on the manuscript, and to Gerardo Franco and
Xiaoxia Xu (both CSM) for useful discussions and suggestions. The work described in our paper has
been carried out during summer internship of I.\ V.\ at Shell~E\,\&\,P in 2005.

\appendix

\section{Small-crack-density approximations of seismic signatures} \label{AppA}

The goal of this Appendix is to develop an intuitive understanding of kinematic seismic signatures
in formations containing multiple sets of vertical fractures. To arrive at tractable expressions,
we assume that the principal crack densities are small, $\{e_1, \, e_2\} \ll 1$, and linearize all
pertinent quantities in $e_1$ and $e_2$. As stated in the main text, arbitrarily oriented fractures
embedded in an otherwise isotropic host rock cause nearly orthorhombic effective symmetry,
therefore, Tsvankin's\cite{Tsvankin1997} anisotropy coefficients introduced for orthorhombic media are of
obvious importance for our study. We begin by deriving these anisotropy coefficients in terms of
the governing parameters $\bm{m}$ (equation~\ref{mfs.eq.50a}) and then turn our attention to
the NMO ellipses of pure modes reflected from horizontal interfaces.

\subsection{Anisotropy coefficients}

As follows from equations~\ref{ru.eq.01}~--~\ref{ru.eq.03}, definition of the effective stiffness
tensor, $\bm{c}_e = \bm{s}_e^{-1}$, and equations 16~--~19, 23~--~26 in Tsvankin\cite{Tsvankin1997}, the
vertical velocities $V_{P0}$, $V_{S0}$ and relevant anisotropy coefficients, fully linearized in
the crack densities, are:
\begin{equation} \label{Vp0}
    V_{P0} = V_{P,\,b} \,
             \left[ 1 + \frac{\, 2 \, \lambda_b^2 \, (\varsigma - 1) \, ({e}_1 + {e}_2 )}
                             {3 \, \mu_b \, (\lambda_b + \mu_b )} \,
             \right] ,
\end{equation}
\begin{equation} \label{Vs0}
    V_{S0} = V_{S,\,b} \left( 1 - \frac{\, 8 \,}{3} \, \, e_1 \,
        \frac{\lambda_b + 2 \, \mu_b}{ 3 \, \lambda_b + 4 \, \mu_b} \right),
\end{equation}
\begin{equation} \label{eps2}
    \epsilon^{(2)} = \frac{\, 8 \,}{3} \, \, {e}_1 \, \left( \varsigma - 1 \right),
\end{equation}
\begin{equation} \label{eps1}
    \epsilon^{(1)} = \frac{\, 8 \,}{3} \, \, {e}_2 \, \left( \varsigma - 1 \right),
\end{equation}
\begin{equation} \label{del2}
    \delta^{(2)} = \frac{\,8\,}{3} \, \, {e}_1
        \left[ \frac{\left( \varsigma - 1 \right) \, \lambda_b}{\lambda_b + \mu_b} -
               \frac{4 \, \mu_b}{3 \, \lambda_b + 4  \, \mu_b} \right] ,
\end{equation}
\begin{equation} \label{del1}
    \delta^{(1)} = \frac{\,8\,}{3} \, \, {e}_2
        \left[ \frac{\left( \varsigma - 1 \right) \, \lambda_b}{\lambda_b + \mu_b} -
               \frac{4 \, \mu_b}{3 \, \lambda_b + 4  \, \mu_b} \right] ,
\end{equation}
\begin{equation} \label{gam2}
    \gamma^{(2)} = - \frac{\,8\,}{3} \, \, {e}_1 \,
        \frac{\lambda_b + 2 \, \mu_b}{3 \, \lambda_b + 4  \, \mu_b} \, ,
\end{equation}
\begin{equation} \label{gam1}
    \gamma^{(1)} = - \frac{\,8\,}{3} \, \, {e}_2 \,
        \frac{\lambda_b + 2 \, \mu_b}{3 \, \lambda_b + 4  \, \mu_b} \, .
\end{equation}
Here
\begin{equation} \label{Vpb}
    V_{P,\,b} = \sqrt{\frac{\lambda_b + 2 \, \mu_b}{\rho}}
\end{equation}
and
\begin{equation} \label{Vsb}
    V_{S,\,b} = \sqrt{\frac{\mu_b}{\rho}}
\end{equation}
are the velocities in the isotropic unfractured rock, $\rho$ is the mass density, and the
quantities $\lambda_b$, $\mu_b$, $e_1$, $e_2$, and $\varsigma$ are described in the main
text. Note that the anisotropy coefficients $\epsilon^{(2)}$, $\delta^{(2)}$, and
$\gamma^{(2)}$, defined in the plane $[{x}_1, \, {x}_3]$ that contains the normal to the
first fracture set with the crack density $e_1$, are proportional to $e_1$ and independent
of $e_2$. Likewise, the coefficients $\epsilon^{(1)}$, $\delta^{(1)}$, and $\gamma^{(1)}$
are controlled by $e_2$ only. 

\subsection{NMO ellipses and ratios of vertical velocities}

Here we present the small-crack-density approximations for the data components
(equation~\ref{mfs.eq.51a}). The ratios of the vertical velocities are
\begin{eqnarray} \label{velrat1}
    \frac{V_{S1}}{V_{P0}} \Bs =
    \frac{V_{S0}}{V_{P0}} \, \sqrt{\frac{1 + 2 \, \gamma^{(1)}}{1 + 2 \, \gamma^{(2)}}} =
    \frac{V_{S,\,b}}{V_{P,\,b}} \,
        \Bigg\{ 1 - \frac{2}{\,3 \, \mu_b \, (\lambda_b + \mu_b)} \,
                \Bigg[ \lambda_b^2 \, \left( \varsigma - 1 \right) \, e_1
    \\
    \Bs \hspace{32mm} + \,
        \frac{(3 \, \lambda_b + 2 \, \mu_b) \,
              (\lambda_b^2 \, \varsigma + 4 \, \mu_b^2) -
              \lambda_b^2 \, (3 \, \lambda_b - 2 \, \mu_b \, \varsigma)}
             {3 \, \lambda_b + 4 \, \mu_b} \, \, e_2 \,
                \Bigg]
        \Bigg\} \,  \qquad \nonumber
\end{eqnarray}
and
\begin{eqnarray} \label{velrat2}
    \frac{V_{S2}}{V_{P0}} \Bs =
    \frac{V_{S0}}{V_{P0}} =
    \frac{V_{S,\,b}}{V_{P,\,b}} \,
        \Bigg\{ 1 - \frac{2}{\,3 \, \mu_b \, (\lambda_b + \mu_b)} \,
                \Bigg[ \lambda_b^2 \, \left( \varsigma - 1 \right) \, e_2
    \\
    \Bs \hspace{32mm} + \,
        \frac{(3 \, \lambda_b + 2 \, \mu_b) \,
              (\lambda_b^2 \, \varsigma + 4 \, \mu_b^2) -
              \lambda_b^2 \, (3 \, \lambda_b - 2 \, \mu_b \, \varsigma)}
             {3 \, \lambda_b + 4 \, \mu_b} \, \, e_1 \,
                \Bigg]
        \Bigg\} \, . \qquad \nonumber
\end{eqnarray}

The pure-mode NMO ellipses from a horizontal reflector beneath a homogeneous orthorhombic
layer are given by\cite{GrechkaTsvankin1998} 
\begin{equation} \label{nmoell1}
    \frac{1}{V_{Q,\,{\rm nmo}}^2(\phi)} =
    W_{Q,\,11} \, \cos^2 \phi + W_{Q,\,22} \, \sin^2 \phi =
    \frac{\cos^2 \phi}{\left[V_{Q,\,{\rm nmo}}^{(1)}\right]^2} +
    \frac{\sin^2 \phi}{\left[V_{Q,\,{\rm nmo}}^{(2)}\right]^2} \, .
\end{equation}
In this equation, $Q =$ P, S$_1$, or S$_2$ is the wave type, $\phi$ is the azimuth measured from the normal to
the first fracture set, $W_{Q,\,11}$ and $W_{Q,\,22}$ are the nonzero elements of the $2 \times 2$
matrices $\bm{W}$ that describe general NMO ellipses, and $V_{Q,\,{\rm nmo}}^{(i)}$ are the
symmetry-direction NMO velocities. These velocities are conveniently expressed in terms of the P-
and S-wave vertical velocities and the anisotropy coefficients listed above\cite{GrechkaTheophanisTsvankin1999}: 
\begin{equation} \label{Vnmop}
    V_{P,\,{\rm nmo}}^{(i)} = V_{P0} \, \sqrt{1 \, + \, 2 \, \delta^{(i)}} \, ,
    \quad (i = 1, \, 2) \, ,
\end{equation}
\begin{equation} \label{Vnmos11}
    V_{S1,\,{\rm nmo}}^{(1)} = V_{S1} \, \sqrt{1 + 2 \, \sigma^{(1)}} \, ,
\end{equation}
\begin{equation} \label{Vnmos12}
    V_{S1,\,{\rm nmo}}^{(2)} = V_{S2,\,{\rm nmo}}^{(1)} =
    V_{S1} \, \sqrt{1 + 2 \, \gamma^{(2)}} =
    V_{S2} \, \sqrt{1 + 2 \, \gamma^{(1)}} \, ,
\end{equation}
\begin{equation} \label{Vnmos22}
    V_{S2,\,{\rm nmo}}^{(2)} = V_{S2} \, \sqrt{1 + 2 \, \sigma^{(2)}} \, ,
\end{equation}
where
\begin{equation} \label{sigma1}
    \sigma^{(1)} = \left( \frac{V_{P0}}{V_{S1}} \right)^2 (\epsilon^{(1)} - \delta^{(1)})
    \, ,
\end{equation}
\begin{equation} \label{sigma2}
    \sigma^{(2)} = \left( \frac{V_{P0}}{V_{S2}} \right)^2 (\epsilon^{(2)} - \delta^{(2)})
    \, .
\end{equation}
Substitution of equations~\ref{Vp0}~--~\ref{Vsb} in equations~\ref{Vnmop}~--~\ref{sigma2}
and subsequent linearization yields
\begin{eqnarray} \label{vpnmo1}
    V_{P,\,{\rm nmo}}^{(1)} \Bs = V_{P,\,b} \,
    \Bigg\{ 1 + \frac{2}{\,3 \, \mu_b \, (\lambda_b + \mu_b)} \,
                \Bigg[ \lambda_b^2 \, \left( \varsigma - 1 \right) \, e_1
    \\
    \Bs \hspace{-5mm} + \,
        \frac{1}{3 \, \lambda_b + 4 \, \mu_b} \,
        \left( 3 \, \lambda_b^3 \, \left( \varsigma - 1 \right)
            + 16 \, \mu_b \, \left( \lambda_b^2 \, \left( \varsigma - 1 \right) +
                    \lambda_b \, \mu_b \, \left( \varsigma - 2 \right) - \mu_b^2 \right)
        \right) e_2 \, \Bigg]
        \Bigg\} \, , \qquad \nonumber
\end{eqnarray}
\begin{eqnarray} \label{vpnmo2}
    V_{P,\,{\rm nmo}}^{(2)} \Bs = V_{P,\,b} \,
    \Bigg\{ 1 + \frac{2}{\,3 \, \mu_b \, (\lambda_b + \mu_b)} \,
                \Bigg[ \lambda_b^2 \, \left( \varsigma - 1 \right) \, e_2
    \\
    \Bs \hspace{-5mm} + \,
        \frac{1}{3 \, \lambda_b + 4 \, \mu_b} \,
        \left( 3 \, \lambda_b^3 \, \left( \varsigma - 1 \right)
            + 16 \, \mu_b \, \left( \lambda_b^2 \, \left( \varsigma - 1 \right) +
                    \lambda_b \, \mu_b \, \left( \varsigma - 2 \right) - \mu_b^2 \right)
        \right) e_1 \, \Bigg]
        \Bigg\} \, , \qquad \nonumber
\end{eqnarray}
\begin{eqnarray} \label{vsnmo11}
    V_{S1,\,{\rm nmo}}^{(1)} = V_{S,\,b} 
    \left\{ 1 + \frac{\,8 \, (\lambda_b + 2 \, \mu_b)}
                     {3 \, (\lambda_b + \mu_b)} \, \, {e}_2 \,
        \left[ \varsigma - \frac{\mu_b}{3 \, \lambda_b + 4 \, \mu_b} \, \right]
    \right\} ,
\end{eqnarray}
\begin{eqnarray} \label{vsnmo12}
    V_{S1,\,{\rm nmo}}^{(2)} = V_{S2,\,{\rm nmo}}^{(1)} = V_{S,\,b} 
    \left[ \, 1 - \frac{8 \, (\lambda_b + 2 \, \mu_b)}
                     {3 \, (3 \, \lambda_b + 4 \, \mu_b)} \, \left( e_1 + {e}_2 \right) \,
    \right] ,
\end{eqnarray}
\begin{eqnarray} \label{vsnmo22}
    V_{S2,\,{\rm nmo}}^{(2)} = V_{S,\,b}
    \left\{ 1 + \frac{\,8\, (\lambda_b + 2 \, \mu_b)}
                     {3 \, (\lambda_b + \mu_b)} \, \, {e}_1 \,
        \left[ \varsigma - \frac{\mu_b}{3 \, \lambda_b + 4 \, \mu_b} \, \right]
    \right\} .
\end{eqnarray}

It is a simple matter now to construct the Fr\'{e}chet-derivative matrix, $\bm{F}$, of the
quantities given by equations~\ref{velrat1}, \ref{velrat2}, and \ref{vpnmo1}~--~\ref{vsnmo22} with
respect to the elements of parameter vector $\bm{m}$ (equation~\ref{mfs.eq.50a}) and verify
that all singular values of $\bm{F}$ are nonzero.

\subsection{Shear-wave splitting coefficient and the eccentricity of P-wave NMO ellipse}

Finally, we use the already derived approximations to demonstrate that such well-known
fracture indicators as the shear-wave splitting coefficient and the eccentricity of P-wave
NMO ellipse are controlled by the {difference} of two principal crack densities, $e_1
- e_2$, and, therefore, insufficient for characterization of multiple fracture sets.

The eccentricity (or normalized elongation) of the P-wave NMO ellipse is just the
difference between two $\delta$ coefficients. As follows from equations~\ref{del2}
and~\ref{del1}, the eccentricity is
\begin{equation} \label{ddel}
    \delta^{(1)} - \delta^{(2)} = \frac{\,8\,}{3} \, \, (e_1 - {e}_2)
        \left[ \frac{\left( 1 - \varsigma \right) \, \lambda_b}{\lambda_b + \mu_b} +
               \frac{4 \, \mu_b}{3 \, \lambda_b + 4  \, \mu_b} \right] \, .
\end{equation}
Similarly, definition of the shear-wave splitting coefficient, $\gamma^{(\rm S)} \approx
\gamma^{(1)} - \gamma^{(2)}$, and equations~\ref{gam2}, \ref{gam1} yield
\begin{equation} \label{gams}
    \gamma^{(\rm S)} = \frac{\,8\,}{3} \, \, (e_1 - {e}_2) \, \,
        \frac{\lambda_b + 2 \, \mu_b}{\, 3 \, \lambda_b + 4  \, \mu_b} \, .
\end{equation}
These signatures unambiguously constrain the crack density only for a single
fracture set, that is, when $e_2 = 0$.



\end{document}